\documentclass{iaus}
\usepackage{graphicx}

\title[Modelling Red Supergiant Populations] 
{Modelling the Near-IR Spectra of Red Supergiant-dominated Populations}

\author[A. Lan\c{c}on et al.]   
{Ariane Lan\c{c}on$^1$, 
 Jay S. Gallagher$^2$, 
 Richard de Grijs$^3$, \break
 Peter Hauschildt$^4$,
 Djazia Ladjal$^5$, 
 Mustapha Mouhcine$^6$, \break 
 Linda J. Smith$^7$, 
 Peter R. Wood$^8$,
 Natascha F\"orster Schreiber$^9$}

\affiliation{$^1$Observatoire de Strasbourg (UMR 7550), 
   11 rue de l'Universit\'e, 67000 Strasbourg, France \break
   email: lancon@astro.u-strasbg.fr\\[\affilskip]
$^2$Dept. of Astronomy, 5534 Sterling, University of Wisconsin, Madison,
   WI 53706, USA \\[\affilskip]
$^3$Dept. of Physics \& Astronomy, University of Sheffield, 
   Hicks Building, Honusfield Rd., Sheffield S3~7RH, UK \\[\affilskip]
$^4$Hamburger Sternwarte, Gojenbergsweg 112, 21029 Hamburg, 
   Germany \\ [\affilskip]
$^5$Institute of Astronomy, Katholieke Universiteit, Celestijnenlaan 200\,B,
 3001 Leuven, Belgium \\[\affilskip]
$^6$Astrophysics Research Institute, Liverpool John Moores University,
 Twelve Quays House, Egerton Wharf, Birkenhead, CH41~1LD, UK \\[\affilskip]
$^7$Space Telescope Science Institute, 3700 San Martin Drive,
   Baltimore, MD 21218, USA \\[\affilskip] 
$^8$RSAA, Mt Stromlo Observatory,
 Cotter Road, Weston Creek, ACT 2611, Australia \\ [\affilskip]
$^9$MPI f\"ur Extraterrestrische Physik, Giessenbachstrasse, 85741 Garching,
   Germany 
 }

\pubyear{2007}
\volume{241}  
\pagerange{1--4}
\date{?? and in revised form ??}
\jname{Stellar Populations as Building Blocks of Galaxies}
\editors{A. Vazdekis et alr., eds.}
\begin{document}

\maketitle

\begin{abstract}
We report on recent progress in the modelling of the 
near-IR spectra of young stellar populations, i.e. populations in which
red supergiants (RSGs) are dominant. First, we discuss the determination
of fundamental parameters of RSGs from {\sc Phoenix} 
model fits to their near-IR spectra; RSG-specific surface 
abundances are accounted for and effects of the microturbulence parameter
are explored.  New population synthesis predictions are then described 
and, as an example, it is shown that the spectra of young star clusters in M\,82 
can be reproduced very well from 0.5 to 2.4\,$\mu$m. 
We warn of remaining uncertainties in cluster ages.  
\keywords{galaxies: stellar content, galaxies: starburst,
galaxies: star clusters, galaxies: individual (M82), infrared: galaxies, 
infrared: stars, stars: supergiants
}
\end{abstract}

\firstsection 
\section{Introduction}

Red supergiant stars (RSGs) provide most of the near-IR light emitted by 
young stellar populations, such as those in starburst galaxies. As 
star forming environments tend to be dusty, rest-frame optical analyses are 
incomplete (highly obscured populations are missed) 
and it is crucial to improve our understanding of 
spectra at longer wavelengths. In the past, the near-IR analysis of 
young stellar populations has often been restricted to the determination of the 
average properties of the dominant stars, such as their spectral
types or abundances.  The subsequent interpretation of these results 
in terms of precise stellar population ages and star formation histories
remains an enormous challenge, as it requires (i) a good
understanding of the near-IR spectra of individual RSGs
and (ii) adequate stellar evolution tracks. 
We have started a programme that aims at 
providing state of the art predictions for the emission of 
RSG-dominated populations and at characterizing remaining
uncertainties. Currently, the project focuses on wavelengths between
0.81 and 2.4\,$\mu$m and spectral resolutions of order $\lambda/\delta 
\lambda = 10^3$. 

\section{Empirical and synthetic spectra of red supergiants}

\begin{figure}[t!]
\includegraphics[clip=,width=0.45\textwidth,height=0.27\textheight]{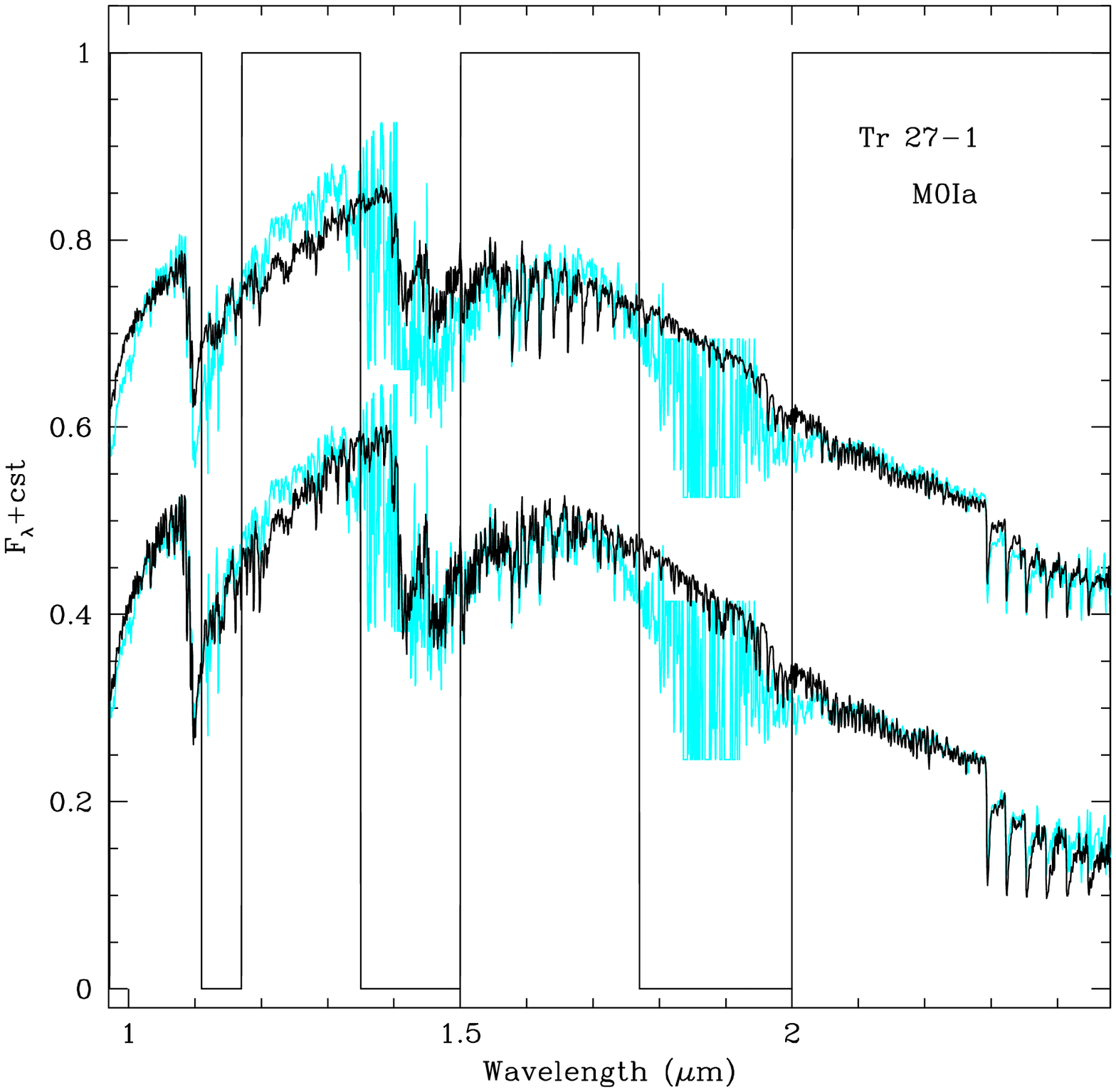}
\includegraphics[clip=,width=0.45\textwidth,height=0.26\textheight]{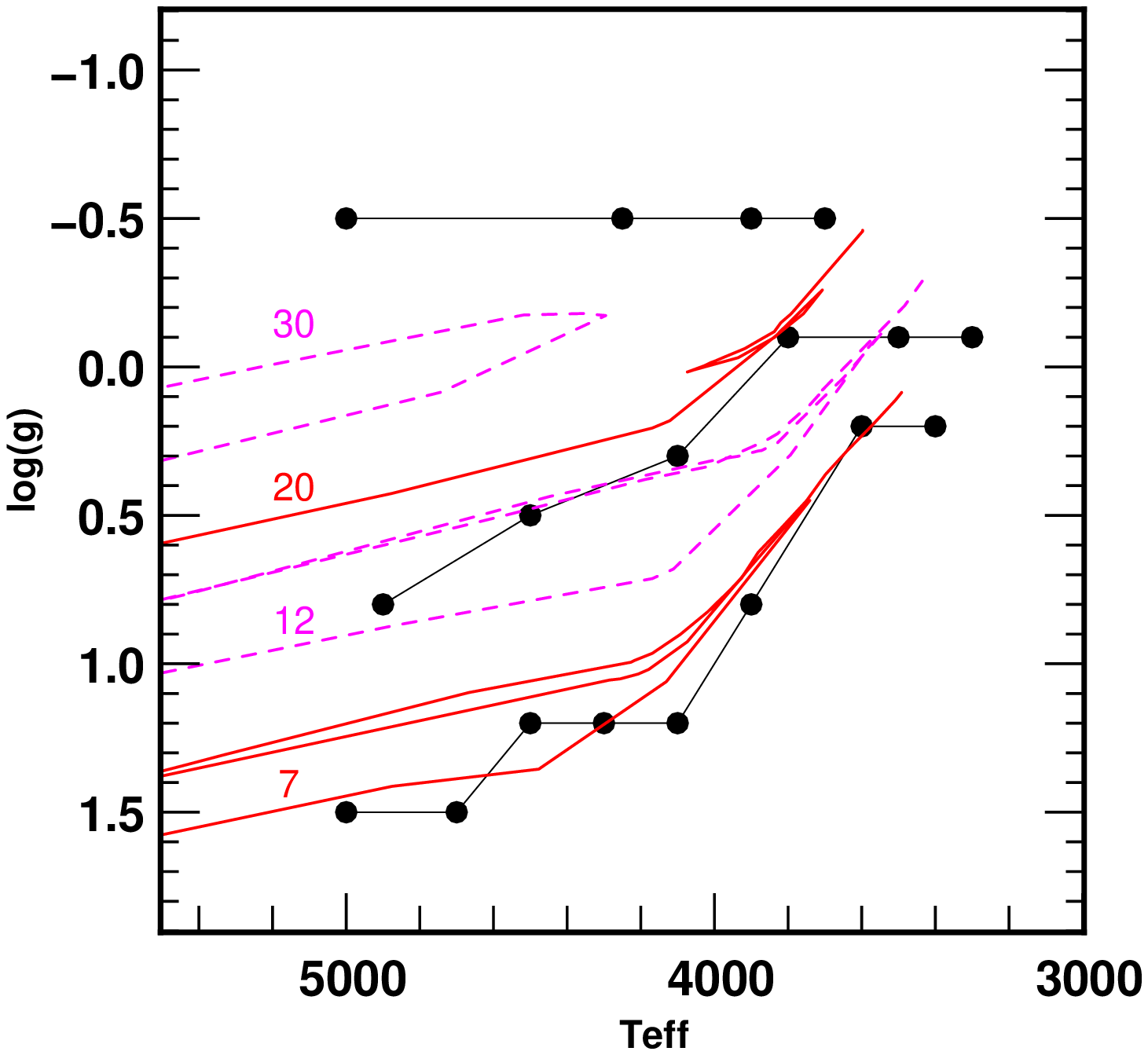}\\
\includegraphics[clip=,width=0.42\textwidth,height=0.24\textheight]{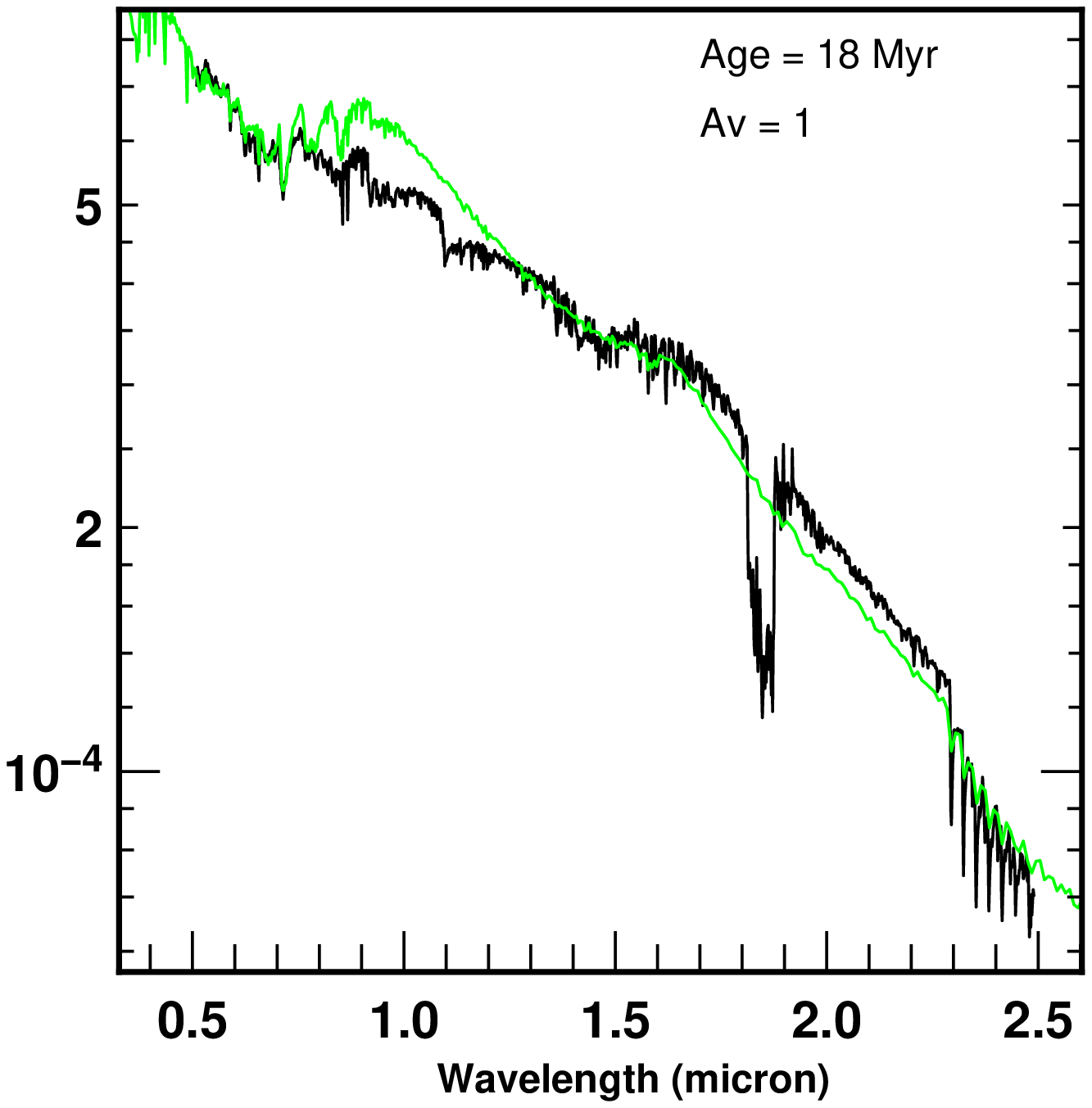}
\hspace{0.02\textwidth}
\includegraphics[clip=,width=0.45\textwidth,height=0.24\textheight]{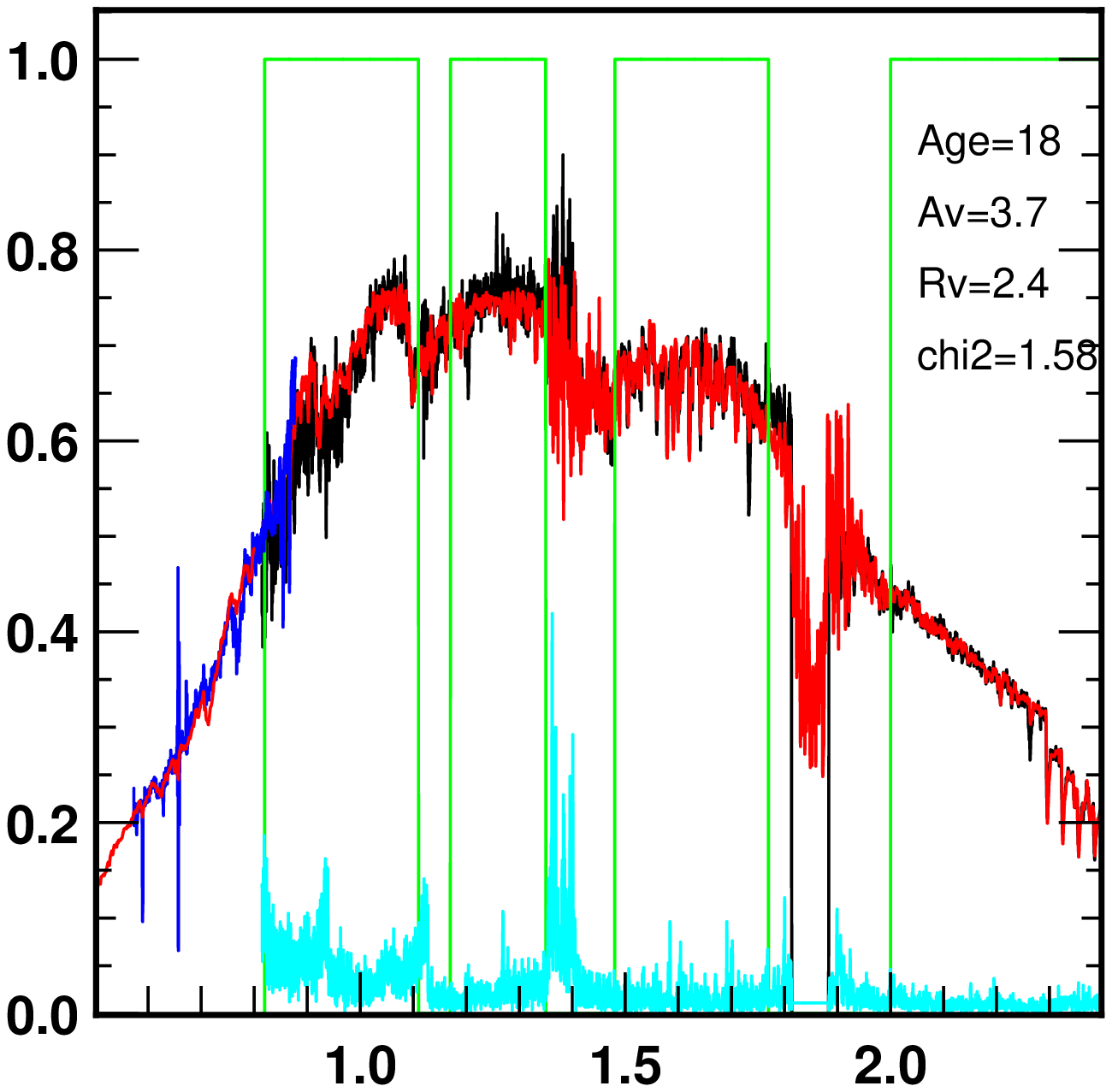}\\
\includegraphics[clip=,width=0.45\textwidth,height=0.27\textheight]{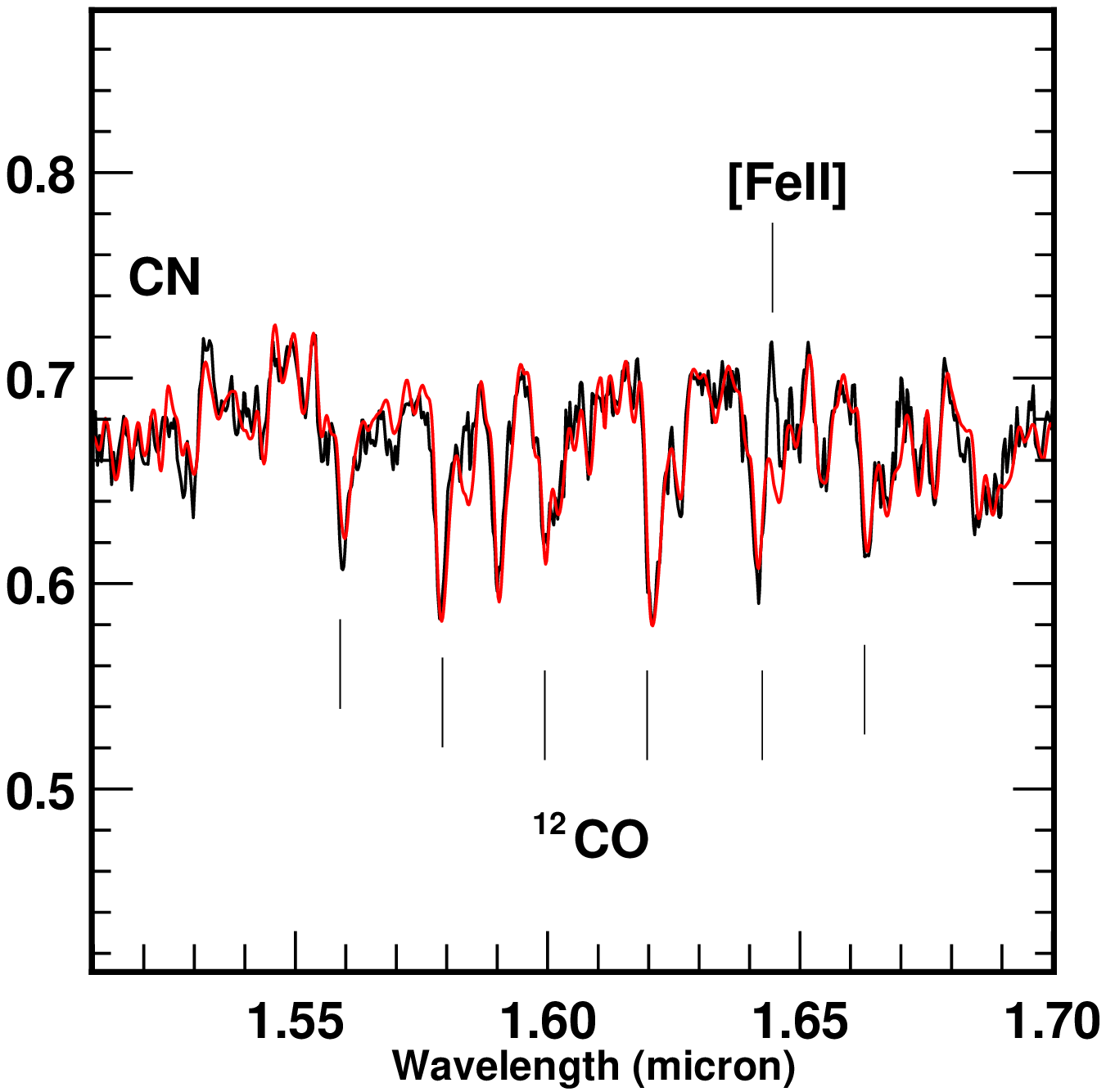}
\includegraphics[clip=,width=0.45\textwidth,height=0.27\textheight]{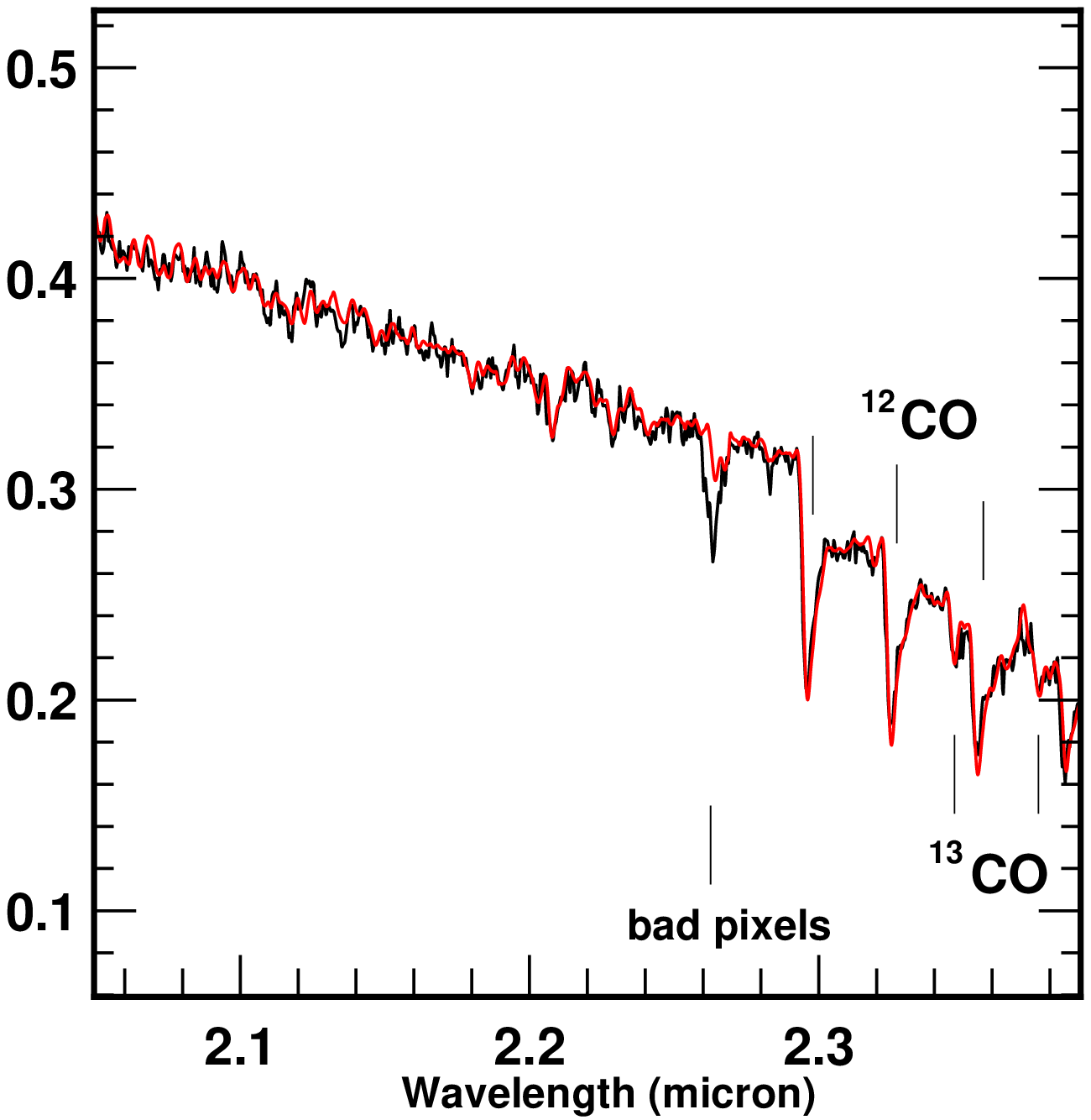}
\caption[]{{\bf Top left:} Spectrum of an M0Ia RSG compared with models
with v$_{\rm microturb}$=2\,km/s (top: 4200\,K, log(g)=-1, A$_V$=4.4) and
with v$_{\rm microturb}$=10\,km/s (bottom: 4500\,K, log(g)=0, A$_V$=4.7; note
the improved CN at 1.1\,$\mu$m and CO around 1.6 and 2.3\,$\mu$m).
{\bf Top right:} Parameters assigned to the new sequences of average spectra, 
superimposed on the solar metallicity tracks of Bressan et al. (1993). 
{\bf Middle left:} Comparison of a new SSP spectrum (black) 
with the standard predictions of Pegase.2 (differences are largest between
10 and 20\,Myr). {\bf Middle right:} Best near-IR fit to the spectrum of
cluster M82-L. The extinction law of Cardelli et al. (1989) with R$_V$=2.4
allows us to also reproduce the optical spectrum (from Smith \&
Gallagher 1999). The error spectrum and the $\chi^2$ weighting function
are shown. {\bf Bottom:} Zoom-ins of the H and K windows.
}
\label{ALfigures.fig}
\end{figure}

In principle, synthetic stellar spectra are more practical for the 
prediction of galaxy spectra than empirical ones, because theory allows 
us to sample parameter space without biases. Lan\c{c}on et al. (2007)
show that modern theoretical spectra can reproduce the near-IR (+optical)
emission of giant stars well down to effective temperatures 
T$_{\rm eff} \simeq 3400$\,K, but that they are not yet satisfactory at lower
temperatures and higher luminosities. They used new {\sc Phoenix} models
to compute spectra at the necessary resolution (0.1\,\AA\ before smoothing),
with solar abundances {\em and} with the RSG-specific abundances 
obtained as the result of internal mixing along stellar evolution tracks; 
the models include some 10$^9$ individual molecular and atomic lines, 
assume spherical symmetry, and allow dust to form if conditions are fulfilled.
Model limitations include the assumptions of local thermal
equilibrium (LTE) and of hydrostatic equilibrium.
A sample of 101 empirical spectra covering wavelengths between 
0.51, 0.81 or 0.90\,$\mu$m and 2.4\,$\mu$m was used for comparison
(Lan\c{c}on \& Wood 2000, Lan\c{c}on et al. in preparation). The 
data were acquired with CASPIR on the 2.3m ANU Telescope at Siding Spring
and with SpeX at IRTF, Hawaii.  Below T$_{\rm eff} \sim 3400$\,K, 
uncertain input line lists are a problem in the models (especially for 
molecular bands around 1\,$\mu$m). At high luminosity (luminosity class Ia
and Iab), the main difficulty is to reproduce 
simultaneously extremely deep CN bands and the relative strengths of the
CO bandheads around 1.7\,$\mu$m and at 2.3\,$\mu$m. RSG-specific abundances
improve fits to the CN bands.  Exploratory calculations show that values
near 10\,km/s for the ``microturbulence" (a 1D-model parameter that hides
poorly understood 3D physical phenomena) may be able to solve both 
problems (Fig.\,\ref{ALfigures.fig}, top left). 
The calculation of a new grid has been launched to explore this
further. In the mean time, the study shows that the population synthesis
community still has to rely on empirical spectra for RSGs, and it warns
that the lack of satisfactory stellar models implies large uncertainties
on the derived fundamental parameters of the observed stars.  

\section{Population synthesis using averaged stellar spectra}

In order to compute spectra of synthetic populations, we have constructed 
three sequences of average empirical spectra, corresponding to luminosity
classes Ia, Iab and Ib/II. Each subset was sorted into bins 
according to the estimated T$_{\rm eff}$, the spectra were dereddened 
(an estimate of the reddening is provided by the model fits), and averages
were computed. The sequences shown in Fig.\,\ref{ALfigures.fig} (top right)
account for varying 
microturbulence in a preliminary way, based on the limited number of 
high microturbulence models available to us at the time of this writing. 
We chose to flag
any star with an initial mass above 7\,M$_{\odot}$ as a supergiant, which
implies that the new spectra affect predictions up to the age of about 75\,Myr
(Fig.\,\ref{ALfigures.fig}, middle left). 
We note that predictions vary significantly depending on the adopted 
evolutionary tracks; different authors predict different red supergiant 
lifetimes, and main sequence rotation affects both
the surface abundances and the final red (and blue) supergiant numbers.

\section{Star clusters in M82}

The synthetic spectra of single stellar populations (SSPs) at solar
metallicity are compared with those of young star clusters in starbursts,
such as M\,82-L and M\,82-F (Smith \& Gallagher 1999). The
selected clusters are massive (well above 10$^5$\,M$_{\odot}$), i.e. 
stochastic effects due to an underpopulated RSG-branch are avoided. A few
have well determined optical ages (based on standard non-rotating
evolutionary tracks). Figure \ref{ALfigures.fig} (middle right and 
bottom) shows cluster L, the cluster observed with SpeX 
with the best signal-to-noise ratio: an excellent fit is obtained 
over the whole available range in wavelength. Such results
make the new models valuable tools 
for purposes such as weak emission line measurements. The $\chi^2$-test
restricted to near-IR wavelengths not affected by strong telluric 
absorption shows that age is formally determined to 
an accuracy of about $\pm 10$\,Myr. Because of strong reddening,
the optical age of cluster L cannot be determined as well as that of cluster F:
50-70\,Myr (Gallagher \& Smith 2001, McCrady et al. 2005, 
Bastian et al. 2007). For F, our current models provide a near-IR
age range of 32 to 46\,Myr. This small but nevertheless significant 
disagreement calls for several comments. (i) Before accounting for 
luminosity-dependent microturbulence, we found a near-IR age of 10 to 20\,Myr;
we hope that our next generation of synthetic stellar spectra will 
significantly reduce uncertainties originating in uncertain fundamental
parameters of stars. (ii) The spectrum used for optical age-dating and our 
near-IR spectrum have different slopes in the region of overlap. This
suggests slightly different positions were observed: 
the obscuration across M82-F is not at all uniform. In addition,
a younger cluster located at very small projected distance 
might contaminate the near-IR data. (iii) Modified stellar tracks 
(e.g. including rotation) might affect optical ages as well as near-IR ones.

\section{Conclusions}
The spectra of young stellar populations at solar metallicity,
observed at R$\sim 10^3$, can now be modelled well from the 
optical through the near-IR. Nevertheless, ages based on near-IR 
spectra remain severely affected by uncertainties. They are due mainly 
to systematic errors, which further work needs to characterize and reduce.  
Errors are associated on one hand with the fundamental parameters of 
red supergiant stars (theoretical spectra, microturbulence, surface 
abundances of C, N and O, non-LTE, variability, winds, giant-supergiant
transition), and on the other with evolutionary tracks (convection, 
opacities, rotation, binarity, effects of a dense environment).
We expect rapid progress in stellar atmosphere models to
provide us with tools to test stellar tracks further.
Complete optical and near-IR spectra of massive clusters 
such as those of M82 are useful test cases for the identification
and correction of systematic errors, but even they are not trivial
to exploit (due to inhomogeneous background populations and extinction, mass
segregation, etc.).

\begin{discussion}

\discuss{Gustafsson}{Do the models with high microturbulence include 
turbulent pressure in a consistent way?}

\discuss{Lan\c{c}on (after discussion with P.H. and H. Lamers)}{No. But
the microturbulent velocities required to reproduce the spectra with 1D models
are supersonic, which suggests that the actual process is not 
microturbulence... Therefore it is unclear how to relate this parameter
of 1D models to pressure.}

\end{discussion}

\end{document}